\documentclass[12pt]{article}

\usepackage{color}
\definecolor{darkblue}{rgb}{0.1,0.1,.7}
\usepackage[colorlinks, linkcolor=darkblue, citecolor=darkblue, urlcolor=darkblue, linktocpage]{hyperref}
\usepackage[]{amsmath}
\usepackage[]{graphicx}
\usepackage[]{latexsym}
\usepackage[utf8]{inputenc}
\usepackage{slashed,graphicx,color,amsmath,amssymb}
\usepackage{mathrsfs}
\usepackage[margin=10pt,font=small,labelfont=bf]{caption}
\usepackage{amscd}
\usepackage{bm}
\usepackage{xcolor}
\usepackage{upgreek}
\usepackage[square, comma, sort&compress,numbers]{natbib}
\usepackage[all,cmtip]{xy}
\usepackage[margin=1.6in]{geometry}
\usepackage{textcomp}
\usepackage[color=cyan!30!white,linecolor=red,textsize=footnotesize]{todonotes}
\numberwithin{equation}{section}
\usepackage[mathscr]{euscript}

\hypersetup{
	pdfstartview={FitH},    
	pdftitle={Modular covariance and uniqueness of JTb deformed CFT},    
	pdfauthor={Aharony, Datta, Giveon, Jiang, Kutasov},     
	colorlinks=true,       
	linkcolor=blue,          
	citecolor=blue!59!black! ,        
	filecolor=magenta,      
	urlcolor=blue           
}

\newcommand{\tr}{\mathrm{Tr}\,}

\newcommand{\ra}{\mathrm{a}}

\newcommand{\rh}{\mathrm{h}}

\newcommand{\rE}{\mathrm{E}}

\newcommand{\rQ}{\mathrm{Q}}

\def\ttb{T{\bar{T}}}
\def\btau{{\bar{\tau}}}

\def\pd{\partial}

\def\bq{{\bar{q}}}

\def\jtbar{J\bar{T}}

\def\eg{{\it e.g.~}}

\def\nn{\nonumber}
\def\pd{\partial}

\def\l1{{{1-loop}}}

\def\n1{\Bigg|_{n=1}}

\def\n{{(n)}}
\def\tr{{Tr}}

\def\vt{\vartheta}

\def\tr{\text{Tr}}
\def\th{\theta}

\def\bq{\bar{q}}

\def\btau{\bar{\tau}}

\def\la{\langle}
\def\ra{\rangle}

\def\rh{{\widehat \mu}}

	\def\pd{\partial}

		\def\jtbar{J\bar{T}}

\def\magic{\mathscr{D}}
\def\serre{\mathbb{D}}

 \makeatletter
 \g@addto@macro\bfseries{\boldmath}
 \makeatother

\begin{document}
\vspace*{-.6in} \thispagestyle{empty}
\vspace{.2in} {\Large
\begin{center}
{\bf Modular covariance and uniqueness \\ of $J\bar T$ deformed CFTs}
\end{center}}
\vspace{.2in}
\begin{center}
{Ofer Aharony$^{1}$, Shouvik Datta$^2$, Amit Giveon$^3$, Yunfeng Jiang$^2$ \& David Kutasov$^4$}
\\
\vspace{.3in}
\small{
$^1$  \textit{Department of Particle Physics and Astrophysics, \\ Weizmann Institute of Science, \\
Rehovot 7610001, Israel.}\\  \vspace{.3cm}
$^2$\textit{Institut f{\"u}r Theoretische Physik,
ETH Z{\"u}rich}},\\
\small{\textit{Wolfgang Pauli Strasse 27,
CH-8093 Z{\"u}rich, Switzerland.}
}
\\ \vspace{.3cm}
$^3$\textit{Racah Institute of Physics, The Hebrew University, \\
Jerusalem 91904, Israel.
}
\\ \vspace{.3cm}
$^4$\textit{EFI and Department of Physics, University of Chicago, \\
5640 S.~Ellis Av., Chicago, IL 60637, USA. \\
}
\vspace{.1in}

\end{center}

\vspace{.2in}

\begin{abstract}
\normalsize{We study families of two dimensional quantum field theories, labeled by a dimensionful parameter $\mu$, that contain a holomorphic conserved $U(1)$ current $J(z)$. We assume that these theories can be consistently defined on a torus, so their partition sum, with a chemical potential for the charge that couples to $J$, is modular covariant. We further require that in these theories, the energy of a state at finite $\mu$ is a function only of $\mu$, and of the energy, momentum and charge of the corresponding state at $\mu=0$, where the theory becomes conformal. We show that under these conditions, the torus partition sum of the theory at $\mu=0$ uniquely determines the partition sum (and thus the spectrum) of the perturbed theory, to all orders in $\mu$, to be that of a $\mu J\bar T$ deformed conformal field theory (CFT). We derive a flow equation for the $J\bar{T}$ deformed partition sum, and use it to study non-perturbative effects. We find  non-perturbative ambiguities for any non-zero value of $\mu$, and comment  on their possible relations to holography.}
\end{abstract}

\vskip 1cm \hspace{0.7cm}

\newpage

\setcounter{page}{1}

\noindent\rule{\textwidth}{.1pt}\vspace{-1.2cm}
\begingroup
\hypersetup{linkcolor=black}
\tableofcontents
\endgroup
\noindent\rule{\textwidth}{.2pt}

\section{Introduction}
\label{sec:1}

In a recent paper \cite{Aharony:2018bad}, we obtained the torus partition sum of a $t T\bar T$ deformed CFT\footnote{See
\eg \cite{Zamolodchikov:2004ce,Smirnov:2016lqw,Cavaglia:2016oda,McGough:2016lol,Giveon:2017nie,Dubovsky:2017cnj,Giveon:2017myj,Shyam:2017znq,Asrat:2017tzd,Giribet:2017imm,Kraus:2018xrn,Cardy:2018sdv,Cottrell:2018skz,Aharony:2018vux,Dubovsky:2018bmo,Taylor:2018xcy,Bonelli:2018kik,Datta:2018thy,Donnelly:2018bef,Babaro:2018cmq,Conti:2018jho,Chakraborty:2018kpr,Chen:2018eqk} for other works.}
from modular invariance, with some qualitative assumptions about the spectrum of the theory. More precisely, we considered a theory with a single scale, associated with a dimensionful coupling $t$, and assumed that the energies of states in that theory, when formulated on a circle of radius $R$, depend only\footnote{Here we mean  dimensionless energies, momenta and coupling, all measured in units of the radius of the circle, $R$.} on $t$ and on the energies and momenta of the corresponding states in the undeformed theory. We showed that, under these assumptions, the torus partition sum of the theory is uniquely determined to all orders in $t$, to be that of the $t T\bar T$ deformation of the theory with $t=0$.

Non-perturbative contributions to the partition sum,  which are due to states whose energies diverge in the limit $t\to 0$, were found to be compatible with modular invariance and finiteness of the partition sum in the limit of zero coupling only for a particular sign of the coupling.\footnote{For this sign, the perturbative spectrum contains states whose energies become complex in the deformed theory, which leads to problems with unitarity.} We discussed possible relations between these field theoretic results and holography.

In this note, we generalize the analysis of \cite{Aharony:2018bad} to the case of a $J\bar T$ deformed CFT. This system was originally discussed in \cite{Guica:2017lia} and the spectrum was obtained in \cite{Chakraborty:2018vja} (see also \cite{Apolo:2018qpq,Bzowski:2018pcy} for other works on this subject). As we will see, the techniques of \cite{Aharony:2018bad} provide a powerful approach  for studying this theory. In particular, we will be able to rederive and extend the results of \cite{Chakraborty:2018vja} using this perspective.

Before turning on the deformation, the current $J$ is holomorphic, i.e. it satisfies $\bar\partial J=0$, as is standard in CFT. As emphasized in \cite{Chakraborty:2018vja}, the $\mu J\bar T$ deformation is essentially defined by the requirement that it preserves this property at arbitrary coupling $\mu$, despite the fact that the full theory is no longer conformal. We will assume in our analysis that this property holds in the theories we discuss.

Usually, in two dimensional field theory, the presence of a holomorphic current means that the theory has an essentially decoupled conformal sector (see \eg \cite{Kutasov:1994xq}). This does not seem to be the case here, probably because the theory is non-local (in the sense that its UV behavior is not governed by a fixed point of the renormalization group). This issue deserves further study.

As in \cite{Aharony:2018bad}, we assume that our theory has a single dimensionful coupling $\mu$. The focus of our discussion is going to be the partition sum of the theory,
\begin{align}\label{eq:ztaunu}
\mathcal{Z}(\tau,\bar{\tau},\nu|\rh)=\sum_n e^{2\pi i\tau_1R P_n-2\pi \tau_2 R \mathcal{E}_n+2\pi i\nu \mathcal{Q}_n},
\end{align}
where $\rh$ is the dimensionless coupling, $\rh\sim \mu/R$, and the sum runs over the eigenstates of the Hamiltonian, the momentum operator $P$, and the charge operator $Q$.  One can think of (\ref{eq:ztaunu}) as the partition sum of the theory on a torus with modulus $\tau=\tau_1+i\tau_2$, in the presence of a chemical potential $\nu$ that couples to the conserved current $J$.

At $\rh=0$, (\ref{eq:ztaunu}) becomes the torus partition sum of a CFT with non-zero chemical potential. It is modular covariant,
\begin{align}
	\label{eq:Ztrans}
	Z_0\left(\frac{a\tau+b}{c\tau+d},\frac{a\bar{\tau}+b}{c\bar{\tau}+d},\frac{\nu}{c\tau+d}\right)
	=\exp\left(\frac{\pi i k c\nu^2}{c\tau+d}\right)Z_0(\tau,\bar{\tau},\nu),
\end{align}
with $a,b,c,d\in\mathbb{Z}$ and $ad-bc=1$. Here $k$ is the level of the $U(1)$ affine Lie algebra,
\begin{align}\label{eq:normJ}
[J_m,J_n] = k \, m \, \delta_{m+n,0}\, .
\end{align}
Note that (\ref{eq:Ztrans}) implies that the chemical potential $\nu$ transforms as a modular form of weight $(-1,0)$. This is due to the fact that it couples to a holomorphic current of dimension $(1,0)$ (see \eg\cite[\textsection \,3.1]{Dijkgraaf:1996iy} for a discussion).

As mentioned above, a key observation of \cite{Chakraborty:2018vja}  was that the current $J$ remains holomorphic in the $J\bar T$ deformed theory as well. Motivated by this, we assume that the partition sum (\ref{eq:ztaunu}) of our theory satisfies a similar modular covariance property,
\begin{align}
	\label{eq:modularZJ}
	\mathcal{Z}\left(\left.\frac{a\tau+b}{c\tau+d},\frac{a\bar{\tau}+b}{c\bar{\tau}+d},\frac{\nu}{c\tau+d}\right|\frac{\rh}{c\bar{\tau}+d}\right)
	=\exp\left(\frac{i\pi k  c\nu^2}{c\tau+d}\right)\mathcal{Z}(\tau,\bar{\tau},\nu|\rh).
\end{align}
The transformation of the (dimensionless) coupling $\rh$ follows from the fact that we assume that it couples in the action to an operator that in the undeformed theory has dimension $(1,2)$\footnote{Notice that for $tT\overline{T}$ deformation, the dimensionful coupling $t$ does not change under modular transformation. Here the dimensionful coupling $\mu=\widehat{\mu}R$ transforms non-trivially under modular transformation. This difference is due to the fact that $\mu$ is not a scalar but has non-zero spin.}. The transformation of $\nu$ is a consequence of the holomorphy of the current $J$, associated with the charge $Q$ that $\nu$ couples to (the discussion of \cite{Dijkgraaf:1996iy} can be extended to this case). Note that in our analysis we take this current to be normalized as in (\ref{eq:normJ}) for all $\rh$. This choice is reflected in the factor of $k$ in the exponential on the right-hand side of (\ref{eq:modularZJ}). It provides the normalization of the charges in (\ref{eq:ztaunu}), which will play an important role in our discussion.

Following the logic of \cite{Aharony:2018bad}, we now ask the following question. Suppose we are given a theory with a single scale, set by a  dimensionful coupling $\mu$, and a current $J(z)$ that is holomorphic throughout the RG flow. Using the fact that the theory on a torus is modular covariant, (\ref{eq:modularZJ}), and assuming that the  energies $\mathcal{E}_n$ and charges $\mathcal{Q}_n$ in (\ref{eq:ztaunu}) depend only on $\rh$ and on the values of the energy, momentum and charge of the corresponding states in the undeformed (conformal) theory, what can we say about the theory?

We will see that, like in \cite{Aharony:2018bad}, the above requirements fix the partition sum (\ref{eq:ztaunu}) uniquely to be that of a $\mu J\bar T$ deformed CFT to all orders in $\rh$. Thus, a $J\bar T$ deformed CFT is the unique theory with these general properties.

In the process of proving that, we will derive equations that govern the flow of energies and charges as a function of the coupling $\rh$. These equations generalize the inviscid Burgers' equation that describes the flow of the energies in a $T\bar T$ deformed CFT \cite{Smirnov:2016lqw,Cavaglia:2016oda}. We will also discuss the theory non-perturbatively in $\rh$ for the $J\bar T$ case\footnote{Note that if we assume that the theory exists at finite $\rh$ and not just in perturbation theory, then this theory seems to be non-local, similar to $T\bar{T}$-deformed theories. In our analysis we do not assume locality, but only the existence of well-defined energy levels of the theory on a circle. This implies that the torus partition sum is well-defined and modular-covariant (for finite $\rh$), without needing to assume locality, and (\ref{eq:modularZJ}) follows.}, by using a differential equation for the partition sum that generalizes the one used in \cite{Cardy:2018sdv,Datta:2018thy}, and discuss relations to holography.

The plan of this paper is the following. In section~\ref{sec:defspectrum} we generalize the discussion of \cite{Aharony:2018bad} to a theory with a holomorphic $U(1)$ current. We show that modular covariance (\ref{eq:modularZJ}), and the qualitative assumption about the spectrum mentioned above, determine the partition sum of the model uniquely to be that of a $\mu J\bar T$ deformed CFT, to all orders in the coupling $\rh$. In particular, we obtain a recursion relation, (\ref{eq:vvv}), satisfied by the partition sum.

In section~\ref{sec:flow-eq} we show that the recursion relation (\ref{eq:vvv}) leads to a flow equation for the partition sum, (\ref{eq:david}), from which one can derive flow equations for the energies and charges of states with the coupling, (\ref{eq:burr}), (\ref{eq:ttburg}), whose solutions agree with the spectrum found in \cite{Chakraborty:2018vja}. We also study the solutions of (\ref{eq:david}) non-perturbatively in $\rh$ and discuss some ambiguities that we find.

In section~\ref{sec:examples} we discuss two examples of our construction -- charged free bosons and fermions. We comment on our results and their relation to holography in section~\ref{sec:conc}.  Two appendices contain results and details that are used in the main text.

\section{Spectrum from modular covariance}
\label{sec:defspectrum}
In this section, we use modular covariance (\ref{eq:modularZJ}), and  the qualitative assumption about the spectrum described in the previous section, to uniquely fix the partition sum to all orders in $\rh$.

We start with the torus partition sum of the theory with $\rh=0$, a CFT with a $U(1)$ current $J$, and a chemical potential $\nu$ for the corresponding charge,
\begin{align}\label{eq:zzerodef}
	Z_0(\tau,\bar{\tau},\nu)=\tr \left[e^{2\pi i \tau(L_0-{c\over 24}) -2\pi i \btau(\bar{L}_0-{c\over 24})+2\pi i\nu J_0}\right] =\sum_n e^{2\pi i\tau_1 R P_n-2\pi \tau_2 R E_n+2\pi i\nu Q_n},
\end{align}
where $P_n$, $E_n$ and $Q_n$ are the momentum, energy and charge of the state $|n\rangle$ on a circle of radius $R$. They are related to the eigenvalues of $L_0,\bar{L}_0$ and $J_0$ by
\begin{align}\label{eq:defcharges}
	\left(L_0-\bar{L}_0\right)|n\rangle=R P_n|n\rangle,\quad\left(L_0+\bar{L}_0-\frac{c}{12}\right)|n\rangle= RE_n|n\rangle,\quad J_0|n\rangle=Q_n|n\rangle.
\end{align}
The partition sum (\ref{eq:zzerodef}) satisfies the modular covariance property (\ref{eq:Ztrans}), which is essentially the statement that the theory can be consistently formulated on a torus.

We now consider a deformation of the CFT, under which the states $|n\rangle_0$ are deformed to $|n\rangle_\rh$, and the quantities in (\ref{eq:defcharges}) become
\begin{align}\label{eq:defpe}
	P_n\mapsto P_n,\qquad E_n\mapsto\mathcal{E}(E_n,P_n,Q_n,\rh),\qquad Q_n\mapsto \mathcal{Q}(E_n,P_n,Q_n,\rh),
\end{align}
where $\rh$ is a dimensionless parameter, which can be thought of as the value of the dimensionful coupling $\mu$ at the scale $R$. The deformation is universal in the sense that the deformed energy and charge of the state $|n\rangle_\rh$ only depend on the values of $(P_n,E_n,Q_n)$ of the undeformed state $|n\rangle_0$.

To evaluate the deformed torus partition sum (\ref{eq:ztaunu}), we follow \cite{Aharony:2018bad} and assume that the quantities in (\ref{eq:defpe}) allow regular Taylor expansions in $\rh$
\begin{align}
	\label{eq:pertEQ}
	&\mathcal{E}_n=\mathcal{E}(E_n,P_n,Q_n,\rh)=\sum_{k=0}^\infty \rE_n^{(k)}\rh^k=\rE_n^{(0)}+\rE_n^{(1)}\rh+\rE_n^{(2)}\rh^2+\cdots,\\\nonumber
	&\mathcal{Q}_n=\mathcal{Q}(E_n,P_n,Q_n,\rh)=\sum_{k=0}^\infty \rQ_n^{(k)}\rh^k=\rQ_n^{(0)}+\rQ_n^{(1)}\rh+\rQ_n^{(2)}\rh^2+\cdots,
\end{align}
where $\rE_n^{(0)}=E_n$ and $\rQ_n^{(0)}=Q_n$, and $\rE_n^{(k)}$,  $\rQ_n^{(k)}$ are functions of $(E_n, P_n, Q_n)$ that need to be determined.

Plugging (\ref{eq:pertEQ}) into (\ref{eq:ztaunu}), we find the Taylor expansion of the deformed partition sum,
\begin{align}
	\label{eq:www}
	\mathcal{Z}(\tau,\bar{\tau},\nu|\rh)=\sum_{p=0}^\infty Z_p \rh^p=Z_0+Z_1\rh+Z_2\rh^2+\cdots.
\end{align}
Modular covariance of the deformed partition sum, (\ref{eq:modularZJ}), implies that $Z_p$ transforms as a non-holomorphic Jacobi form of weight $(0,p)$ and holomorphic index $k$,
\begin{align}\label{eq:formppp}
	Z_p\left(\frac{a\tau+b}{c\tau+d},\frac{a\bar{\tau}+b}{c\bar{\tau}+d},\frac{\nu}{c\tau+b}\right)=(c\bar{\tau}+d)^p\,\exp\left(\frac{i \pi  k c\nu^2}{c\tau+d}\right)\,Z_p(\tau,\bar{\tau},\nu).
\end{align}
The first few orders in the $\rh$ expansion are given by
\begin{align}
	\label{eq:expZp}
	Z_p=\sum_{n}f_n^{(p)}e^{2\pi i\tau_1 R P_n-2\pi \tau_2 R E_n+2\pi i\nu Q_n},
\end{align}
where
\begin{align}
	\label{eq:expfn}
	f_n^{(1)}=&\,(-2\pi R\rE_n^{(1)})\tau_2+2i\pi\nu\rQ_n^{(1)},\\\nonumber
	f_n^{(2)}=&\,\frac{1}{2!}\left(-2\pi R\rE_n^{(1)}\right)^2\tau_2^2-2\pi R\left[\rE_n^{(2)}+2i\pi\nu\rE_n^{(1)}\rQ_n^{(1)}\right]\tau_2
	-2\left[\pi^2\nu^2(\rQ_n^{(1)})^2-i\pi\nu\rQ_n^{(2)}\right],\\\nonumber
	f_n^{(3)}=&\,\frac{1}{3!}\left(-2\pi R\rE_n^{(1)}\right)^3\tau_2^3+4\pi^2R^2\left[\rE_n^{(1)}\rE_n^{(2)}+i\pi\nu(\rE_n^{(1)})^2\rQ_n^{(1)}\right]\tau_2^2\\\nonumber
	&\,+\left[2R\pi^2\nu^2\rE_n^{(1)}(\rQ_n^{(1)})^2-2\pi i R\nu(\rE_n^{(1)}\rQ_n^{(2)}+\rE_n^{(2)}\rQ_n^{(1)})-2\pi R\rE_n^{(3)}\right]\tau_2\\\nonumber
	&\,-\frac{4}{3}i\pi^3\nu^3(\rQ_n^{(1)})^3-4\pi^2\nu^2\rQ_n^{(1)}\rQ_n^{(2)}+2\pi i\nu\rQ_n^{(3)}.
\end{align}
As in \cite{Aharony:2018bad}, we can write $Z_p$ as a differential operator in $\tau$, $\nu$ acting on $Z_0$, by replacing  $\rE_n^{(k)}(E_n,P_n,Q_n)$, $\rQ_n^{(k)}(E_n,P_n,Q_n)$ in (\ref{eq:expfn})  by differential operators, using the  replacement rules
\begin{align}
	E_n\mapsto -\frac{1}{2\pi R}\partial_{\tau_2},\qquad P_n\mapsto \frac{1}{2\pi iR}\partial_{\tau_1},\qquad Q_n\mapsto\frac{1}{2\pi i}\partial_{\nu}.
\end{align}
This leads to a double expansion of $Z_p$ in powers of $\tau_2$ and $\nu$,
\begin{align}
	\label{eq:expandzp}
	Z_p=\sum_{l,m}\tau_2^l\nu^m {\mathcal{O}}_{lm}^{(p)}(\partial_{\tau},\partial_{\bar{\tau}},\partial_\nu)Z_0,
\end{align}
where the sum runs over the range $l,m=0,1,\cdots, p$; $0<l+m\le p$.

As is clear from the expansion (\ref{eq:ztaunu}), (\ref{eq:pertEQ}),  the differential operators $ {\mathcal{O}}_{lm}^{(p)}(\partial_{\tau},\partial_{\bar{\tau}},\partial_\nu)$ with given $p$ are only sensitive to the energy and charge shifts $\rE_n^{(k)}$,  $\rQ_n^{(k)}$ with $k=1,2,\cdots, p$. Conversely, if we know all $ {\mathcal{O}}_{lm}^{(p)}$ with given $p$, we can determine all the energy and charge shifts with $k\le p$ by using (\ref{eq:expZp}) -- (\ref{eq:expandzp}).

We can use the expansion  (\ref{eq:expandzp}) to prove that if $Z_1,\cdots, Z_p$ have been determined, $Z_{p+1}$ can be determined as well. As in \cite{Aharony:2018bad}, we start by considering the first step in this process.  Equation (\ref{eq:expandzp}) (with $p=1$) takes in this case the form
\begin{align}\label{eq:zone}
	Z_1=\left(\tau_2\widehat{O}^{(1)}_{1,0}(\partial_{\tau},\partial_{\bar{\tau}},\partial_\nu)
	+\nu\widehat{O}^{(1)}_{0,1}(\partial_{\tau},\partial_{\bar{\tau}},\partial_\nu)\right)Z_0.
\end{align}
We are looking for differential operators $\widehat{O}^{(1)}_{1,0}$, $\widehat{O}^{(1)}_{0,1}$, for which $Z_1$ transforms as a Jacobi form of weight $(0,1)$ and index $k$, for any $Z_0$ of weight $(0,0)$ and index $k$. To find them, one can proceed as follows.

In \cite{Datta:2018thy,Aharony:2018bad}, we used the modular covariant derivative operators
\begin{align}
\label{eq:defdd}
\textsf{D}_{\tau}^{(r)}\equiv\partial_{\tau}-\frac{ir}{2\tau_2},\qquad \textsf{D}_{\bar{\tau}}^{(\bar r)}\equiv\partial_{\bar{\tau}}+\frac{i\bar r}{2\tau_2}.
\end{align}
These operators have the following properties. Acting with $\textsf{D}_{\tau}^{(r)}$ on a modular form of weight $(r,\bar r)$ gives a modular form of weight $(r+2, \bar r)$. Similarly, $\textsf{D}_{\bar{\tau}}^{(\bar r)}$ increases the weight of such a modular form to $(r, \bar r+2)$.\footnote{Acting with $\textsf{D}_{\bar{\tau}}^{(\bar r)}$ on a \emph{Jacobi form} of weight $(r,\bar{r})$ and holomorphic index $k$ gives a Jacobi form of weight $(r,\bar{r}+2)$ with the same index. On the other hand, acting with $\textsf{D}_{{\tau}}^{( r)}$ on a Jacobi form with holomorphic index $k\not=0$ does not give a Jacobi form.}

In our case, it is useful to introduce another covariant derivative, with respect to $\nu$,
\begin{align}
	\label{eq:Dnu}
	\textsf{D}_\nu\equiv\partial_{\nu}+\frac{\pi k \nu}{\tau_2}.
\end{align}
Acting with $\textsf{D}_\nu$ on a Jacobi form of weight $(r,\bar r)$ and index $k$ gives a Jacobi form of weight $(r+1,\bar r)$ and index $k$ (see appendix\,\ref{sec:covD} for more details).

Using the covariant derivatives in  (\ref{eq:defdd}), (\ref{eq:Dnu}), it is straightforward to find a combination of the form (\ref{eq:zone}) that has the correct modular transformation properties,
\begin{align}\label{eq:first-ord}
	Z_1=\alpha\tau_2\textsf{D}_\nu\partial_{\bar{\tau}}Z_0.
\end{align}
Here $\alpha$ is a constant that can be absorbed in the definition of $\rh$;  we will set it to one below. It is not hard to check that (\ref{eq:first-ord})  is the unique object of the form (\ref{eq:zone}) with the correct modular transformation properties.

We are now ready to move on to the general induction step. We assume that $Z_1,\cdots, Z_p$ (with $p\ge 1$) have been determined, and want to show that $Z_{p+1}$ can be determined as well.

We saw before that from the form of $Z_1,\cdots, Z_p$ we can read off the energy and charge shifts $\rE_n^{(k)}$,  $\rQ_n^{(k)}$ with $k=1,2,\cdots, p$. Consider now the expansion (\ref{eq:expandzp}) of $Z_{p+1}$. Most of the terms in that expansion involve the energy and charge shifts with $k\le p$, which are assumed to be already known. There are only two terms in the sum, corresponding to $(l,m)=(1,0)$ and $(0,1)$, that involve the unknowns $\rE_n^{(p+1)}$,  $\rQ_n^{(p+1)}$.

To show that there is no more than one solution for the expansion (\ref{eq:expandzp}), suppose there were two different ones. Subtracting them, and using the fact that most terms in the expansion (\ref{eq:expandzp}) cancel between the two, we find that there must exist differential operators $\delta\widehat{O}^{(p+1)}_{1,0}(\partial_{\tau},\partial_{\bar{\tau}},\partial_\nu)$, $\delta\widehat{ O}^{(p+1)}_{0,1}(\partial_{\tau},\partial_{\bar{\tau}},\partial_\nu)$, such that
\begin{align}
	\left(\tau_2\,\delta\widehat{O}^{(p+1)}_{1,0}(\partial_\tau,\partial_{\bar{\tau}},\partial_\nu)
	+\nu\,\delta\widehat{ O}^{(p+1)}_{0,1}(\partial_\tau,\partial_{\bar{\tau}},\partial_\nu)\right)Z_0
\end{align}
is a Jacobi form of weight $(0,p+1)$ and index $k$, for any $Z_0$ which is a Jacobi form of weight (0,0) and index $k$. The fact that such differential operators do not exist (for $p>0$) can be proven by using the properties of the covariant derivatives (\ref{eq:defdd}), (\ref{eq:Dnu}), in a similar way to the proof for the $T\bar T$ case in \cite{Aharony:2018bad}, and we will not  repeat it here.

So far, we have proved that if a $Z_p$ with the right properties exists, it is unique. In order to prove existence, one can again proceed as in the $T\bar T$ case \cite{Aharony:2018bad}.  There, it followed from a recursion relation that gave $Z_{p+1}$ in terms of $Z_p$. It is natural to seek a similar recursion relation in our case. It turns out that such a relation exists, but it is more complicated. In particular, it relates $Z_p$ to all $Z_j$ with $0\le j<p$. It takes the form
\begin{align}
	\label{eq:vvv}
	Z_{p}=\frac{\tau_2}{p}\left[\textsf{D}_\nu\textsf{D}_{\bar{\tau}}^{(p-1)}-\frac{i\pi k \nu(p-1)}{2\tau_2^2}\right]Z_{p-1}
	-\frac{i\pi k}{2p}\sum_{j=0}^{p-2}\left(\frac{\pi\nu k }{2i\tau_2}\right)^j\textsf{D}_{\bar{\tau}}^{(p-j-2)}Z_{p-j-2}~.
\end{align}
One way to arrive at this recursion relation is to start with the known spectrum of the theory \cite{Chakraborty:2018vja}, plug it into the partition sum (\ref{eq:ztaunu}), and expand in $\rh$. Alternatively, $Z_p$ can be determined order by order by taking an ansatz consisting of terms with  the appropriate modular properties and demanding it has the general form (\ref{eq:expandzp}). The structure of this expansion at low $p$ is discussed  in appendix~\ref{sec:Zpcov}.

In the next section we will prove that (\ref{eq:vvv}) indeed provides a solution of (\ref{eq:expandzp}) for all $p$, which establishes that under the assumptions we described above, the partition sum (\ref{eq:ztaunu}) is uniquely determined to all orders in $\rh$.

Our discussion of uniqueness in this section started from the assumption that the coupling $\mu$ has dimension $(0,-1)$, i.e. that the corresponding perturbing operator has dimension $(1,2)$. More generally, if $\mu$ has dimension $(h,\bar h)$, i.e. the corresponding perturbing operator has dimension $(1-h,1-\bar h)$, the dimensionless coupling $\rh$ transforms under the modular group as a form of weight $(h,\bar h)$, and $Z_1$ transforms as a Jacobi form of weight $-(h,\bar h)$ and index $k$. One can show that the form (\ref{eq:zone}) is inconsistent with this transformation property, except for the case $h=0$, $\bar h=-1$ that was analyzed above.

\section{Non-perturbative analysis}
\label{sec:flow-eq}

The recursion relation (\ref{eq:vvv}) can be phrased as a differential equation for the partition sum (\ref{eq:ztaunu}).
Namely, if the partition sum $\mathcal{Z}(\tau,\bar{\tau},\nu|\rh)$ satisfies
\begin{align}
	\label{eq:david}
	\left(1+\frac{i\pi k \rh\nu}{2\tau_2}\right)\partial_\rh\mathcal{Z}=\tau_2\textsf{D}_\nu\mathcal{D}_{\bar{\tau}}\mathcal{Z}
	-\frac{i\pi k \rh}{2}\frac{1}{1+\frac{i\pi k \rh\nu}{2\tau_2}}\mathcal{D}_{\bar{\tau}}\mathcal{Z},
\end{align}
where (compare to (\ref{eq:defdd}))
\begin{align}
	\label{eq:amit}
	\mathcal{D}_{\bar{\tau}}\equiv\partial_{\bar{\tau}}+{i\over 2\tau_2}\rh\partial_\rh,
\end{align}
then expanding this equation in a power series in $\rh$ reproduces (\ref{eq:vvv}). For the $\ttb$ case, the flow equation for the torus partition sum can also be derived from a description with a dynamical metric \cite{Cardy:2018sdv,Dubovsky:2018bmo}. It would be interesting to derive \eqref{eq:david} from a similar point of view, by including a dynamical gauge field as well.

As in \cite{Aharony:2018bad}, although (\ref{eq:david}) was derived from a perturbative expansion in $\rh$, we assume that it holds non-perturbatively as well.
Before turning to a discussion of the non-perturbative effects implied by (\ref{eq:david}), we would like to point out that from this equation we can read off a system of differential equations that describes the evolution of the energies and momenta of states with the coupling $\rh$. To do that, we plug the general expression for the partition sum (\ref{eq:ztaunu}) into (\ref{eq:david}), and compare the coefficients of a given exponential on the left and right hand sides.
We also multiply by the factor $(1+i\pi k \widehat{\mu}\nu/(2\tau_2))$ on both sides. The resulting equation then takes the form
\begin{align}
\mathcal{Y}_0 + \mathcal{Y}_1 \nu + \mathcal{Y}_2 \nu^2 =0.
\end{align}
Here, $\mathcal{Y}_i$ are functions containing $\mathcal{E}_n(\rh),\,\mathcal{Q}_n(\rh),\,   {P}_n$ and the derivatives $\mathcal{E}'_n(\rh),\,\mathcal{Q}'_n(\rh)$. Since this should hold for all values of $\nu$, we have $\mathcal{Y}_{0,1,2}=0$. The equations $\mathcal{Y}_1- \mathcal{Y}_2=0$ and  $\mathcal{Y}_2=0$ respectively yield
\begin{align}\label{eq:burr}
&\mathbb{E}'_n(\rh)\left[1+\pi\rh\mathcal{Q}_n(\rh)\right]=\pi\left[\mathbb{P}_n-\mathbb{E}_n(\rh)\right]\mathcal{Q}_n(\rh),\\\nonumber
&\mathcal{Q}'_n(\rh)\left[1+\pi\rh\mathcal{Q}_n(\rh)\right]=\frac{\pi {k}}{2}\left[\mathbb{P}_n-\mathbb{E}_n(\rh)\right],
\end{align}
where $\mathbb{E}_n(\rh)=R\mathcal{E}_n(\rh)$, and $\mathbb{P}_n=RP_n$ is the quantized momentum. The equation $\mathcal{Y}_0=0$ gives rise to a equation which is consistent with the above two.


Dividing the two equations in (\ref{eq:burr}), one finds that
\begin{align}\label{eq:coset}
k\mathbb{E}_n(\rh)-\mathcal{Q}_n(\rh)^2={\rm independent\;of\;\rh,}
\end{align}
which reproduces one of the results of \cite{Chakraborty:2018vja}.

Equations (\ref{eq:burr}) can be expressed in a form that is closer to Burgers' equation by writing them in terms of the dimensionful $\mu J\bar T$ coupling $\mu=\rh R$, and using the fact that the dimensionless energies $\mathbb{E}_n$ depend only on the dimensionless coupling $\rh$.

The resulting system of equations can be written as\footnote{To reproduce the equations given in \cite{Chakraborty:2018vja}, we need to make the replacement $\rh=\mu/(2\pi R)$.}
\begin{align}\label{eq:ttburg}
	&\frac{\partial}{\partial \mu} (\mathcal{E}_n-P_n)=\,\pi\mathcal{Q}_n\frac{\partial}{\partial R} (\mathcal{E}_n-P_n),\\\nonumber
	&\frac{\partial \mathcal{Q}_n}{\partial \mu}=\,\pi\mathcal{Q}_n\frac{\partial \mathcal{Q}_n}{\partial R}-\frac{\pi {k}}{2}(\mathcal{E}_n-P_n).
\end{align}
The differential equation on the second line of (\ref{eq:ttburg}) looks like the inviscid Burgers' equation with a time-dependent source, where the coupling $\mu$ plays the role of time. The dynamics of this source is described by the first line of (\ref{eq:ttburg}).

The solution of (\ref{eq:ttburg}) with the boundary conditions $\mathcal{E}_n(0)=E_n$ and $\mathcal{Q}_n(0)=Q_n$ is given by
\begin{align}\label{eq:enqn}
	\mathcal{E}_n^{(+)}(\rh)=&\,-\frac{2}{\pi^2\rh^2 {k}R}\sqrt{(1+\pi Q_n\rh)^2+\pi^2\rh^2 {k}R(P_n-E_n)}\\\nonumber
	&\,+\frac{1}{\pi^2\rh^2{k} R}\left(2+2\pi Q_n\rh+\pi^2\rh^2 {k} P_n R\right),\\\nonumber
	\mathcal{Q}_n^{(+)}(\rh)=&\,\frac{1}{\pi\rh}\sqrt{(1+\pi Q_n\rh)^2+\pi^2\rh^2{k}R(P_n-E_n)}-\frac{1}{\pi\rh},
\end{align}
where we took the positive branch of the square root, so that
\begin{align}\label{eq:limeq}
\lim_{\rh\to 0} \mathcal{E}_n^{(+)}(\rh)=E_n,\qquad \lim_{\rh\to 0} \mathcal{Q}_n^{(+)}(\rh)=Q_n.
\end{align}
Plugging (\ref{eq:enqn}) into (\ref{eq:ztaunu}) gives a partition sum that has a regular Taylor expansion in $\rh$, and satisfies $\lim_{\rh\to0} \mathcal{Z}(\tau,\bar{\tau},\nu|\rh)=Z_0(\tau,\bar{\tau},\nu)$ (\ref{eq:zzerodef}).

It is instructive to compare the spectrum of energies and momenta described by (\ref{eq:enqn}) to that obtained in the $t T\bar T$ case \cite{Smirnov:2016lqw,Cavaglia:2016oda}. There, the structure of the spectrum was different for the two different signs of the coupling $t$. For positive $t$, the energies were real and the asymptotic high energy density of states exhibited Hagedorn growth \cite{Giveon:2017nie,Datta:2018thy}. For $t<0$, states with sufficiently high energy in the original CFT had the property that their energies became complex in the deformed theory.

In the $\mu J\bar T$ case, the spectrum (\ref{eq:enqn}) is the same for both signs\footnote{We take $\rh$ to be real so that the Lagrangian of the theory is real in Lorentzian signature. This is related to the fact that for complex $\rh$, the energies and charges (\ref{eq:enqn}) are in general complex.} of $\rh$, which are related by the symmetry $J\to -J$. The spectrum has the qualitative structure of that with $t<0$ in the $t T\bar T$ case. Beyond a certain maximal undeformed (right-moving) energy, that depends on the charge, the deformed energy and charge become complex. As in the $t T\bar T$ case, this means that the theory is not unitary; the consequences of this remain to be understood.

It is also worthwhile to note that, as in the $\ttb$ case \cite{Datta:2018thy}, the spectrum \eqref{eq:enqn} contains a protected subsector. States with $E_n=P_n$ retain their CFT charges and energies in the $J\bar T$ deformed theory. This is natural from the perspective of (\ref{eq:burr}); it is related to the fact that $E_n-P_n$ is the charge that couples to $\bar T$, which appears in the interaction Lagrangian. If the original CFT has a right-moving supersymmetry, states with $E_n=P_n$ are right-moving Ramond ground states, and the spectrum \eqref{eq:enqn} implies that the elliptic genus with a chemical potential for $Q$ does not depend on $\rh$.

Note also that the discussion of this section provides a proof of the statement that the recursion relation (\ref{eq:vvv}) gives rise to a solution of (\ref{eq:expandzp}) for all $p$. Indeed, this recursion relation is equivalent to the flow equations  (\ref{eq:david}) -- (\ref{eq:burr}), which give rise to the spectrum (\ref{eq:enqn}). Plugging this spectrum into (\ref{eq:ztaunu}) gives $Z_p$'s of the form (\ref{eq:expandzp}).

We now move on to a discussion of non-perturbative contributions to the partition sum that solves (\ref{eq:david}). As explained in \cite{Aharony:2018bad}, a simple way to investigate them is to consider the contribution to the partition sum of states for which we take the negative branch of the square root in (\ref{eq:enqn}). The two branches are related by
\begin{align}
	\mathcal{E}_n^{(+)}+\mathcal{E}_n^{(-)}=&\,\frac{2}{\pi^2\rh^2 {k} R}\left(2+2\pi Q_n\rh+\pi^2\rh^2 {k} P_n R\right),\\\nonumber
	\mathcal{Q}_n^{(+)}+\mathcal{Q}_n^{(-)}=&\,-\frac{2}{\pi\rh}.
\end{align}
While $\mathcal{E}_n^{(+)}$, $\mathcal{Q}_n^{(+)}$ approach finite limits as $\rh\to 0$, (\ref{eq:limeq}), $\mathcal{E}_n^{(-)}$, $\mathcal{Q}_n^{(-)}$ diverge in this limit,
\begin{align}\label{eq:limeqneg}
\mathcal{E}_n^{(-)}(\rh)\simeq \frac{4}{\pi^2\rh^2 {k} R},\qquad \mathcal{Q}_n^{(-)}(\rh)\simeq -\frac{2}{\pi\rh}.
\end{align}
The fact that the energy $\mathcal{E}_n^{(-)}$ goes to $+\infty$ in the limit, implies that states with these energies give non-perturbative contributions to the partition sum, which satisfy the correct boundary conditions $\lim_{\rh\to0} \mathcal{Z}(\tau,\bar{\tau},\nu|\rh)=Z_0(\tau,\bar{\tau},\nu)$, as in the $t T\bar T$ case with $t<0$.

One way to find consistent non-perturbative contributions is then to assume that we have some extra states in our theory labeled by ${\tilde n}$, whose energies and charges are given by ${\mathcal{E}}_{\tilde n}^{(-)}$ and ${\mathcal{Q}}_{\tilde n}^{(-)}$ (appearing in the partition sum (\ref{eq:ztaunu})). These states can be the negative branch energies and charges of some other $J\bar T$ deformed CFT, that a priori need not have anything to do with the one that gives the perturbative contributions discussed above.

We find
\begin{align}\label{eq:zznnpp}
	\mathcal{Z}_{\text{np}}=&\,\sum_{\tilde n} e^{2\pi i\tau_1RP_{\tilde n}-2\pi\tau_2 R {\mathcal{E}}_{\tilde n}^{(-)}+2\pi i\nu {\mathcal{Q}}_{\tilde n}^{(-)}}\\\nonumber
	=&\,e^{-\frac{8\tau_2}{\pi {k}\rh^2}-\frac{4i\nu}{\rh}}\sum_{\tilde n} e^{2\pi i\tau_1RP_{\tilde n}+2\pi\tau_2 R {\mathcal{E}}_{\tilde n}^{(+)}-2\pi i\nu {\mathcal{Q}}_{\tilde n}^{(+)}-8\tau_2 Q_{\tilde n}/\rh-4\pi R\tau_2 P_{\tilde n}}.
\end{align}
Using the relation
\begin{align}
	Q_n=\frac{\pi\rh {k} R}{2}\left({\mathcal{E}}_n^{(\pm)}(\rh)-P_n\right)+{\mathcal{Q}}_n^{(\pm)}(\rh),
\end{align}
satisfied by both branches of (\ref{eq:enqn}), we can rewrite (\ref{eq:zznnpp}) as
\begin{align}
	\label{eq:Znp2}
	\mathcal{Z}_{\text{np}}=e^{\frac{\pi{k}\tilde{\nu}^2}{2\tau_2}-\frac{\pi{k}\nu^2}{2\tau_2}}\sum_{\tilde n} e^{2\pi i\tau_1RP_{\tilde n}-2\pi\tau_2 R{\mathcal{E}}_{\tilde n}^{(+)}+2\pi i\tilde{\nu}{\mathcal{Q}}_{\tilde n}^{(+)}},
\end{align}
where the shifted chemical potential is given by
\begin{align}\label{eq:nnt}
	\tilde{\nu}=-\nu+\frac{4 i\tau_2}{\pi{k}\rh}~.
\end{align}
A few comments are in order here:
\begin{enumerate}
\item By construction, the partition sum (\ref{eq:Znp2}) must be modular invariant (since the original expression (\ref{eq:zznnpp}) is). This can be shown directly as follows. The prefactors in (\ref{eq:Znp2}) transform as
\begin{align}
	\label{eq:two-phases}
	e^{-\frac{\pi{k}\nu^2}{2\tau_2}}\mapsto e^{-\frac{\pi {k}\nu^2}{2\tau_2}}\times e^{\frac{ic{k}\pi\nu^2}{c\tau+d}},\qquad
	e^{+\frac{\pi{k}\tilde{\nu}^2}{2\tau_2}}\mapsto e^{+\frac{\pi{k}\tilde{\nu}^2}{2\tau_2}}\times e^{-\frac{ic\pi{k}\tilde{\nu}^2}{c\tau+d}}.
\end{align}
The partition sum on the right-hand side of (\ref{eq:Znp2}) transforms as
\begin{align}
	\sum_{\tilde n} e^{2\pi i\tau_1P_{\tilde n}-2\pi\tau_2 R{\mathcal{E}}_{\tilde n}^{(+)}+2\pi i\tilde{\nu}{\mathcal{Q}}_{\tilde n}^{(+)}}
	\mapsto e^{\frac{ic{k}\pi\tilde{\nu}^2}{c\tau+d}}
	\sum_{\tilde n} e^{2\pi i\tau_1P_{\tilde n}-2\pi\tau_2 R{\mathcal{E}}_{\tilde n}^{(+)}+2\pi i\tilde{\nu}{\mathcal{Q}}_{\tilde n}^{(+)}}.
\end{align}
Combining these transformations, we see that $\mathcal{Z}_{\text{np}}$ (\ref{eq:Znp2}) indeed transforms as a Jacobi form,
(\ref{eq:modularZJ}).
\item
The fact that $\mathcal{Z}_{\text{np}}$ is a non-perturbative contribution to the partition sum is due to the behavior as $\rh\to 0$ of the prefactor on the right-hand side of (\ref{eq:Znp2}). The leading behavior of the partition sum in this limit is $\mathcal{Z}_{\text{np}}\sim e^{-\frac{8\tau_2}{\pi k\rh^2}}\tilde Z_0$, which is exponentially small for both signs of $\rh$, as expected. Thus, we see that the non-perturbative completion of the partition sum of $J\bar T$ deformed CFT has a similar ambiguity to that found in \cite{Aharony:2018bad} for a $t T\bar T$ deformed CFT with negative $t$. This was perhaps to be expected, since already the perturbative spectrum of the theory showed a similar structure (complex energies) to that encountered in that case.
\item In the analysis of this section, the case $i\pi k \rh\nu+2\tau_2=0$ plays a special role. In particular, eq. (\ref{eq:david}) degenerates in that case, and (\ref{eq:nnt}) takes the form $\tilde{\nu}=\nu$. Furthermore, the partition sum (\ref{eq:ztaunu}) in this case takes the form
\begin{align}
\mathcal{Z}=\mathrm{Tr}\left[ e^{2\pi i\tau(L_0-\bar L_0)+2\pi i\nu J_0}\right].
\end{align}
The above trace is highly divergent, since there is no suppression of states with large $L_0+\bar L_0$ and fixed $L_0-\bar L_0$. Thus, the result depends on the order in which the sum is performed. It would be interesting to understand the physical interpretation of these observations better.
\end{enumerate}

\def\cR{\mathscr{R}}

\section{Examples}
\label{sec:examples}
In this section, we illustrate the discussion of the previous sections by considering two examples, namely, the charged free boson and fermion.

\subsection{Charged free boson}
Consider the CFT of a scalar field $X$, living on a circle of radius $\mathscr{R}$, $X\sim X+2\pi \cR$.
We take the holomorphic current $J$ that figures in the discussion of the previous sections to be $J=i\partial X$. Taking $X$ to be canonically normalized, $\la X(z)X(w)\ra =-\log |z-w|^2$, corresponds to setting the level $k$, (\ref{eq:normJ}), to one. The charge $Q$ is in this case the left-moving momentum, $Q=p_L$.

The partition sum (\ref{eq:zzerodef}) takes the form
\begin{align}
	Z(\tau,\btau,\nu)= \tr \left[q^{L_0-c/24}{\bq}^{\bar{L}_0-c/24} y^{J_0}\right] = \frac{1}{|\eta(\tau)|^2} \sum_{m,n \in \mathbb{Z}} q^{p_L^2\over2}\bq^{p_R^2\over2} y^{p_L},
\end{align}
where $y=e^{2\pi i \nu}$. The left and right-moving momenta are given by
\begin{align}
	p_L = \frac{n}{\cR} + \frac{m \cR}{2} \qquad , \qquad p_R = \frac{n}{\cR} - \frac{m \cR}{2}.
\end{align}
In the decompactification limit, $\cR\to \infty$,  one finds
\begin{align}\label{eq:charged-bose-pf}
	Z(\tau,\btau,\nu) =  \frac{\cR}{|\eta(\tau)|^2} \int_{-\infty}^{\infty} dp \,  (q\bar q)^{p^2\over 2} y^p
	= \frac{\cR}{\sqrt{2\tau_2}\, |\eta(\tau)|^2} \  {\exp\left[-\frac{\pi\nu^2}{2\tau_2} \frac{}{}\right]}.
\end{align}
In the above, $\sqrt{\tau_2}\, |\eta(\tau)|^2$ is invariant under modular transformations. The anomalous transformation factor of the Jacobi form arises from the exponential factor.

\par
The Lagrangian for the $\jtbar$ deformed free boson theory was computed in \cite{Chakraborty:2018vja}.
We consider the expansion for the deformed partition sum \eqref{eq:www}.
The first order correction  is given by \eqref{eq:first-ord}.
Substituting $Z_0$ from \eqref{eq:charged-bose-pf}, we have
\begin{align}
	Z_1 &=  \frac{\pi i \nu}{2\tau_2} Z_0 .
\end{align}
The above quantity transforms as a Jacobi form of  weight (0,1) and holomorphic index 1, as expected. Interestingly, all higher order corrections turn out to vanish
and the $\rh$-expansion terminates,
leading to the following closed form expression for the deformed partition sum:
\begin{align}
\label{eq:def-Z-bos}
	\mathcal{Z}(\tau,\btau,\nu|\rh)
	&= \frac{\cR}{\sqrt{2\tau_2}\, |\eta(\tau)|^2} \  \left(1+\frac{\pi i \rh  \nu}{2\tau_2} \right)\ {\exp\left[-\frac{\pi\nu^2}{2\tau_2} \right]}.
\end{align}
A few comments about this result:
\begin{enumerate}
\item
One can check that \eqref{eq:def-Z-bos} is an exact solution of the flow equation \eqref{eq:david}.
\item \eqref{eq:def-Z-bos} vanishes when $i\pi \rh \nu +2\tau_2 =0$. Note that this is the same value as that discussed in point (3) in the previous section.
\item Any function $\mathcal{F}(\tau,\bar{\tau},\nu|\rh)$ of the form
	\begin{align}
	\mathcal{F}(\tau,\btau,\nu|\rh)
	&= F(\tau,\btau) \  \left(1+\frac{\pi i \rh  \nu}{2\tau_2} \right)\ {\exp\left[-\frac{\pi}{2}\frac{\nu^2}{\tau_2} \right]},
	\end{align}
for an arbitrary function $F(\tau,\bar{\tau})$, is a solution to the flow equation \eqref{eq:david}.\footnote{This expression with an arbitrary $F(\tau,\bar{\tau})$ is a solution to the flow equation \eqref{eq:david}. However, for the full partition sum to have the appropriate modular properties \eqref{eq:modularZJ} we require $F(\tau,\btau)$ to be modular invariant.}  It would be interesting to understand this freedom better.
\end{enumerate}

\subsection{Charged free fermion}

Here we consider a free complex left-moving fermion $(\psi,\psi^*)$ and its right-moving counterpart $(\bar\psi,\bar\psi^*)$. The central charge of the model is $c_L=c_R=1$. The holomorphic current $J$ is given in this case by $J=\psi^*\psi$. Normalizing the fermions canonically, $\la\psi^*(z)\psi(w)\ra=1/(z-w)$, leads to $k=1$ in (\ref{eq:normJ}).

After summing over spin structures, the charged partition sum takes the form
\begin{align}
	{Z}(\tau,\btau,\nu) = \sum_{i=2,3,4}  {\vartheta_i(\nu|\tau) \over \eta(\tau) }{\vartheta_i(0|\btau) \over \eta(\btau)} .
\end{align}
Using the S-modular transformation of the Jacobi $\vartheta$-functions we have
\begin{align}
	Z(\tau,\btau,\nu) = e^{- \frac{\pi i \nu^2}{\tau}}Z(-1/\tau,-1/\btau,\nu/\tau).
\end{align}

We next consider the $\rh$ expansion of the deformed partition sum \eqref{eq:www}.
The first order correction from \eqref{eq:first-ord} is
\begin{align}
	Z_1 =  \tau_2 \sum_i  \bigg({\mathbb{D}^{(1/2)}_{\btau } \vartheta_i(0|\btau) \over \eta(\btau)}\bigg) \bigg({\mathscr{D}^{(0)}_{\nu } \vartheta_i(\nu|\tau)\over \eta(\tau)}\bigg) .
\end{align}
This has  weight (0,1) and holomorphic index 1.
Here, $\mathbb{D}^{(r)}_{\btau }$ is the Ramanujan-Serre derivative \cite{Zagier} which preserves holomorphy and raises the weight $r$ of a modular form by two units
\begin{align}
	\mathbb{D}^{(r)}_{\btau } \equiv \pd_{\btau}- \frac{\pi i r}{6} E_2 (\btau),
\end{align}
and $\mathscr{D}_\nu^{(n)}$ is given in \eqref{eq:curly-D-def}.

The expressions for higher order corrections get progressively more complicated. However, they can be expressed in terms of covariant modular derivatives which simplifies them to some extent. This also facilitates an easy way to read off the modular weights and indices. The second order correction is
\begin{align}
	Z_2=\sum_i& \Bigg[\frac{\tau_2}{6}{\magic_\nu ^{(1)} \vt_i(\nu|\tau) \over    \eta(\tau)}{ \left( 3\tau_2 \serre^{(5/2)}_{\btau}\serre^{(1/2)}_{\tau} \vt_i(0|\btau) +i \pi \tau_2 \tilde{E}_2 (\btau) \nn  \serre^{(1/2)}_{\tau}\vt_i(0|\btau)\right)   \over    \eta(\btau)}\\
	&\ \ - {    i \pi       \over 2}
	{      \left(\nu \magic^{(0)}_\nu \th_i (\nu|\tau)+\th_i (\nu|\tau)\right)   \over   \eta(\tau)} { \serre^{(1/2)}_{\tau} \vt_i(0|\btau)\over \eta(\btau)} \Bigg].
\end{align}
As expected, the above quantity transforms as a Jacobi form of weight (0,2) and holomorphic index 1. In the above formula, the shifted Eisenstein series $\tilde{E}_2 (\btau) $  is a non-holomorphic modular form of weight (0,2), defined as
$
	\tilde{E}_2 (\btau) \equiv {E}_2 (\btau) +{3}/{(\pi\tau_2)}
$
\cite{Zagier}.

\section{Discussion}
\label{sec:conc}
In the recent paper \cite{Aharony:2018bad} we showed that modular invariance together with a qualitative assumption about the spectrum of a two dimensional QFT determine uniquely the partition sum (and thus the spectrum) of the theory to be that of a $T\bar T$ deformed CFT. The main purpose of this note was to generalize the discussion to the case where the QFT contains a holomorphic $U(1)$ current $J$ throughout its RG flow, and the qualitative assumption involves the $U(1)$ charges.

We showed that if such a theory can be defined on a torus, so that its partition sum with a chemical potential for the charge $Q$ associated with $J$ is modular covariant, \eqref{eq:modularZJ}, and it has the further property that the energies and charges of states in the deformed theory depend only on the coupling, $\rh$, and on the spectrum of the undeformed theory, the partition sum and thus the spectrum of energies and charges of the deformed theory is uniquely determined to be that of a $\mu J\bar T$ deformed CFT, to all orders in $\rh \sim \mu/R$.

In the process, we derived a flow equation that governs the evolution of the partition sum with the coupling $\rh$, \eqref{eq:david}, and flow equations that determine the evolution of the energies and charges of states with $\rh$, \eqref{eq:ttburg}, whose solution \eqref{eq:enqn} agrees with that obtained by other means in \cite{Chakraborty:2018vja}.

Studying the flow equation \eqref{eq:david} non-perturbatively, we found ambiguities corresponding to the contributions to the partition sum of states whose energies diverge as the coupling $\rh\to 0$.

In the $t T\bar T$ case, the properties of the theory were found to be sensitive to the sign of the coupling. For one sign ($t>0$ in \cite{Aharony:2018bad}), the energies of all states are real (on a large circle), and the entropy interpolates between the Cardy entropy of a CFT and a Hagedorn entropy. For $t<0$, the energies of highly excited states are complex in the deformed theory, leading to problems with unitarity. Non-perturbatively, the theory with $t>0$ is well defined, while that with $t<0$ has non-perturbative ambiguities.

In $\mu J\bar T$ deformed CFTs, the structure we found for both signs of the coupling $\rh$ is similar to that of a $t T\bar T$ deformed CFT with negative $t$. The energies of highly excited states \eqref{eq:enqn} are complex, and non-perturbatively there are ambiguities. It is an interesting challenge to understand all these theories better. Note that truncating the theory to keep only the real energies is not consistent with modular invariance (and, thus, with a well-defined theory on a torus). In the $T\bar T$ case a specific suggestion for the UV completion of the theory, in terms of a theory of Jackiw-Teitelboim gravity, appeared in \cite{Dubovsky:2017cnj,Dubovsky:2018bmo}; it would be interesting to find a similar UV completion for the $J\bar T$ deformed CFTs.

One of the motivations for studying $J\bar T$ deformed CFT's comes from holography. As in the $T\bar T$ case, there are two different holographic constructions that were considered in the literature. One is the double trace deformation (identical to the large central charge limit of the deformation we discuss in this paper),
that was discussed in \cite{Bzowski:2018pcy};
the conjectured dual geometry is $AdS_3$ with a modification of the boundary conditions
that involves a combination of the metric and of the Chern-Simons gauge field dual to the $U(1)$ current.\footnote{The results in \cite{Bzowski:2018pcy} do not precisely agree with our results for the shifted energy levels and charges, but a small change in the precise boundary conditions, that are in principle determined by the form of the double-trace deformation, should cure this.}
As in the $T \bar T$ case \cite{Aharony:2018bad}, it would be interesting to understand the status of states with complex energies, and the non-perturbative ambiguities that we found, from the bulk point of view. We leave this for future work.

The second holographic construction, discussed in \cite{Chakraborty:2018vja,Apolo:2018qpq},
corresponds to adding to the Lagrangian of the CFT a certain dimension $(1,2)$ single trace operator, $A(x,\bar x)$, constructed in \cite{Kutasov:1999xu}. This operator has the quantum numbers of $J\bar T$, but as explained in \cite{Kutasov:1999xu} it is different from it.  As reviewed in \eg\cite{Chakraborty:2018vja} (see also \cite{Aharony:2018bad}),
for some purposes one can think about the boundary CFT corresponding to string theory on $AdS_3$ as a symmetric product $M^N/S_N$ \cite{Argurio:2000tb,Giveon:2005mi}.
From the point of view of this theory, the operator $A$ takes the form $\sum_{i=1}^N (J\bar T)_i$, and the single trace deformation discussed in \cite{Chakraborty:2018vja,Apolo:2018qpq}
takes the symmetric product to $M_\rh^N/S_N$, where $M_\rh$ is a $J\bar T$ deformed version of the block $M$.

From the bulk point of view, the single trace deformation takes $AdS_3\times S^1$ to a four dimensional background that was described in \cite{Chakraborty:2018vja,Apolo:2018qpq}.\footnote{The authors of \cite{Apolo:2018qpq} studied a concrete example, in which the $S^1$ is embedded in an $S^3$.}
This background is non-singular, but it has closed timelike curves at large values of the radial coordinate, starting from a radial position that depends on $|\rh|$ \cite{Song:2011sr}. Dimensionally reducing it to three dimensions gives rise to null warped $AdS_3$ \cite{Israel:2004vv,Detournay:2005fz,Azeyanagi:2012zd},
a background that plays a role in various developments related to the Kerr/CFT correspondence,
three dimensional Schr\"odinger spacetimes, and dipole backgrounds
(see \eg\cite{Compere:2012jk,ElShowk:2011cm} and references therein for reviews).

It is interesting to compare the properties of the boundary and bulk theories in the $T\bar T$ and $J\bar T$ cases. As discussed above, on the field theory side, many properties of $J\bar T$ deformed CFTs are analogous to those of a $t T\bar T$ deformed CFT with negative $t$. In particular, the energy spectrum becomes complex in the UV, and the partition sum has non-perturbative ambiguities.

On the bulk side with single trace deformations, in the $t T\bar T$ case the background has a curvature singularity at a finite value of the radial coordinate, and closed timelike curves beyond it. In the $J\bar T$ case, there is no curvature singularity, but there are closed timelike curves at large values of the radial coordinate \cite{Song:2011sr}. Thus, it is natural to conjecture that the complex energies and non-perturbative ambiguities mentioned above are related to the closed timelike curves and not to the curvature singularity.

It would be interesting to understand this relation better, and in particular understand whether the theory is well defined after all, despite the issues with unitarity, non-perturbative ambiguities and closed timelike curves.
One possible way to go about this is to further explore the string theory formulation of the theory, as a current-current deformation of string theory on $AdS_3\times S^1$ \cite{Chakraborty:2018vja,Apolo:2018qpq}. We leave this too for future work.

\section*{Acknowledgements}
We thank S. Yankielowicz for discussions. The work of OA and AG was supported in part  by the I-CORE program of the Planning and Budgeting Committee and the Israel Science Foundation (grant number 1937/12) and by an Israel Science Foundation center for excellence grant (grant number 1989/14). The work of OA was also supported by the Minerva foundation with funding from the Federal German Ministry for Education and Research. OA is the Samuel Sebba Professorial Chair of Pure and Applied Physics. The work of SD and YJ is supported by the NCCR SwissMAP, funded by the Swiss National Science Foundation.
The work of DK is supported in part by DOE grant DE-SC0009924. DK thanks the Hebrew University, Tel Aviv University and  the Weizmann Institute for hospitality during part of this work.

\appendix
\section{A covariant derivative}
\label{sec:covD}
In this appendix, we show that the differential operator
\begin{align}
	\textsf{D}_\nu\equiv\partial_\nu+\frac{k\pi\nu}{\tau_2}
\end{align}
is modular covariant. Acting with $\textsf{D}_\nu$ on a non-holomorphic Jacobi form of weight $(n,\bar{n})$ and holomorphic index $k$ gives a Jacobi form of weight $(n+1,\bar{n})$ and holomorphic index $k$.\par

Let us consider a Jacobi form $J_{n,\bar{n}}(\tau,\bar{\tau},\nu)$ of weight $(n,\bar{n})$ and index $k$. Under modular transformations, we have
\begin{align}
	\label{eq:Jnn}
	J_{n,\bar{n}}\left(\tau',\bar{\tau}',\nu'\right)
	=e^{\frac{i\pi ck\nu^2}{c\tau+d}}(c\tau+d)^n(c\bar{\tau}+d)^{\bar{n}}\,J_{n,\bar{n}}(\tau,\bar{\tau},\nu),
\end{align}
where
\begin{align}
	\tau'=\frac{a\tau+b}{c\tau+d},\qquad\bar{\tau}'=\frac{a\bar{\tau}+b}{c\bar{\tau}+d},\qquad\nu'=\frac{\nu}{c\tau+d}.
\end{align}
Acting with $\partial_{\nu}$ on both sides of (\ref{eq:Jnn}) and using the fact that $\partial_{\nu'}=(c\tau+d)\partial_{\nu}$, we have
\begin{align}
	\label{eq:dnuJ}
	\partial_{\nu'}J_{n,\bar{n}}\left(\tau',\bar{\tau}',\nu'\right)=&\,e^{\frac{i\pi ck\nu^2}{c\tau+d}}(c\tau+d)^{n+1}(c\bar{\tau}+d)^{\bar{n}}\,\partial_{\nu}J_{n,\bar{n}}(\tau,\bar{\tau},\nu)\\\nonumber
	&+{2\pi i c k \nu}\,e^{\frac{i\pi ck\nu^2}{c\tau+d}}(c\tau+d)^{n}(c\bar{\tau}+d)^{\bar{n}}\,J_{n,\bar{n}}(\tau,\bar{\tau},\nu).
\end{align}
Multiplying  both sides of (\ref{eq:Jnn}) by $k\pi\nu/\tau_2$, and using the fact that $\nu=\nu' (c\tau+d)$ as well as $\tau_2=\tau'_2(c\tau+d)(c\bar{\tau}+d)$, we find that
\begin{align}
	\label{eq:pinuJ}
	\frac{k\pi\nu'}{\tau'_2}J_{n,\bar{n}}\left(\tau',\bar{\tau}',\nu'\right)
	=&\,(c\bar{\tau}+d)\frac{k\pi\nu}{\tau_2}e^{\frac{i\pi ck\nu^2}{c\tau+d}}(c\tau+d)^{n}(c\bar{\tau}+d)^{\bar{n}}\,J_{n,\bar{n}}(\tau,\bar{\tau},\nu)\\\nonumber
	=&\,\frac{k\pi\nu}{\tau_2}e^{\frac{i\pi ck\nu^2}{c\tau+d}}(c\tau+d)^{n+1}(c\bar{\tau}+d)^{\bar{n}}\,J_{n,\bar{n}}(\tau,\bar{\tau},\nu)\\\nonumber
	&\,-{2\pi i c k \nu}\,e^{\frac{i\pi ck\nu^2}{c\tau+d}}(c\tau+d)^{n}(c\bar{\tau}+d)^{\bar{n}}\,J_{n,\bar{n}}(\tau,\bar{\tau},\nu),
\end{align}
where in the second equality we used the fact that
$
	c\tau+d=c\bar{\tau}+d-2ic\tau_2.
$
Taking the sum of (\ref{eq:dnuJ}) and (\ref{eq:pinuJ}), we see that the terms that are not covariant cancel, and we are left with
\begin{align}
	\textsf{D}_{\nu'}J_{n,\bar{n}}\left(\tau',\bar{\tau}',\nu'\right)
	=e^{\frac{i\pi ck\nu^2}{c\tau+d}}(c\tau+d)^{n+1}(c\bar{\tau}+d)^{\bar{n}}\,\textsf{D}_{\nu}J_{n,\bar{n}}(\tau,\bar{\tau},\nu),
\end{align}
which means that $\textsf{D}_{\nu}J_{n,\bar{n}}(\tau,\bar{\tau},\nu)$ is a Jacobi form of weight $(n+1,\bar{n})$ and index $k$. This completes the proof.

\section{$Z_p$ from a covariant ansatz}
\label{sec:Zpcov}
In order to find $Z_p$, we write down an ansatz with the desired modular properties, (\ref{eq:formppp}), and require it to be consistent with the general structure of the perturbative expansion, (\ref{eq:expandzp}). The leading term in $1/\tau_2$  is fixed by (\ref{eq:expandzp}), (\ref{eq:first-ord}), to be
\begin{align}
	Z_p=\frac{\tau_2^p}{p!}\mathscr{D}_{\nu}^{(p)}\mathscr{D}_{\bar{\tau}}^{(p)}Z_0+\cdots,
\end{align}
where
\begin{align}\label{eq:curly-D-def}
	\mathscr{D}_{\nu}^{(j)}\equiv\textsf{D}_{\nu}^j,\qquad \mathscr{D}_{\bar{\tau}}^{(j)}\equiv\prod_{m=0}^{j-1}\textsf{D}_{\bar{\tau}}^{(2m)}.
\end{align}
The other terms have lower powers of $\tau_2$ and can be written in terms of $\mathscr{D}_{\nu}^{(i)}$, $\mathscr{D}_{\bar{\tau}}^{(j)}$ with $0\le i,j\le p$. A term of the form $\mathscr{D}_{\nu}^{(i)}\mathscr{D}_{\bar{\tau}}^{(j)}Z_0$ with particular $i,j$ is multiplied by $\tau_2^a\nu^b$, such that its contribution to $Z_p$ transforms as a Jacobi form of weight $(0,p)$ and index $k$, (\ref{eq:formppp}). Since $\tau_2^a\nu^b\mathscr{D}_{\nu}^{(i)}\mathscr{D}_{\bar{\tau}}^{(j)}$ has weight
\begin{align}
	(i-a-b,2j-a),
\end{align}
we have the constraint
\begin{align}
	i-a-b=0,\qquad 2j-a=p.
\end{align}
The indices $i,j$ satisfy the constraits $0\le i,j\le p$ and $0\le b\le p$. This leads to
\begin{align}
	0\le p+i-2j\le p.
\end{align}
In addition, there are no terms with $a=b=0$. Taking into account these constraints, we can write down the ansatz for any $p$. The first few $Z_p$ take the form
\begin{align}
	Z_1=&\,a_1\,\tau_2\mathscr{D}_{\nu}^{(1)}\mathscr{D}_{\bar{\tau}}^{(1)}Z_0,\\\nonumber
	Z_2=&\,\left(b_4\,\tau_2^2\mathscr{D}_{\nu}^{(2)}\mathscr{D}_{\bar{\tau}}^{(2)}+b_3\,\nu^2\mathscr{D}_{\nu}^{(2)}\mathscr{D}_{\bar{\tau}}^{(1)}  +b_2\,\nu\mathscr{D}_{\nu}^{(1)}\mathscr{D}_{\bar{\tau}}^{(1)}+b_1\,\mathscr{D}_{\bar{\tau}}^{(1)}\right)Z_0,\\\nonumber
	Z_3=&\,\left(c_7\,\tau_2^3\mathscr{D}_{\nu}^{(3)}\mathscr{D}_{\bar{\tau}}^{(3)}+c_6\,\tau_2\nu^2\mathscr{D}_\nu^{(3)}\mathscr{D}_{\bar{\tau}}^{(2)}  +c_5\,\tau_2\nu\mathscr{D}_\nu^{(2)}\mathscr{D}_{\bar{\tau}}^{(2)}+c_4\,\tau_2\mathscr{D}_{\nu}^{(1)}\mathscr{D}_{\bar{\tau}}^{(2)}\right)Z_0\\\nonumber
	&\,+\frac{1}{\tau_2}\left(c_3\,\nu^3\mathscr{D}_{\nu}^{(2)}\mathscr{D}_{\bar{\tau}}^{(1)}+c_2\,\nu^2\mathscr{D}_{\nu}^{(1)}\mathscr{D}_{\bar{\tau}}^{(1)}  +c_1\,\nu\mathscr{D}_{\bar{\tau}}^{(1)}\right)Z_0
\end{align}
To fix the constants $a_k, b_k, c_k$ we impose the conditions which stem from the structure of the perturbative expansion. To be more explicit, we first expand the covariant derivatives $\mathscr{D}_{\nu}^{(j)}$ and $\mathscr{D}_{\bar{\tau}}^{(j)}$ in terms of $\partial_{\nu}$ and $\partial_{\bar{\tau}}$ in the ansatz. Comparing with the structure of the perturbative expansion, we impose the following conditions
\begin{itemize}
	\item The coefficient of $\tau_2^p\partial_{\nu}^{p}\partial_{\bar{\tau}}^{p}$ is fixed to be $1/p!$;
	\item The coefficients of the terms without $\tau_2$ and $\nu$, namely $\partial_{\nu}^n\partial_{\bar{\tau}}^m$ are zero;
	\item The coefficients of terms with negative powers of $\tau_2$, i.e. terms of the form $\tau_2^{-n}\nu^m\partial_{\nu}^{i}\partial_{\bar{\tau}}^{j}$ with $n>0$, vanish.
\end{itemize}
We find that these conditions are powerful enough to fix $Z_p$ completely at any given order. 
The solutions for the first few orders are given by
\begin{align}
	Z_1=&\,\left(\tau_2\mathscr{D}_{\nu}^{(1)}\mathscr{D}_{\bar{\tau}}^{(1)}\right)Z_0,\\\nonumber
	Z_2=&\,\left(\frac{1}{2}\tau_2^2\mathscr{D}_{\nu}^{(2)}\mathscr{D}_{\bar{\tau}}^{(2)} -\frac{i\pi}{2}\nu\mathscr{D}_{\nu}^{(1)}\mathscr{D}_{\bar{\tau}}^{(1)}-\frac{i\pi}{2}\mathscr{D}_{\bar{\tau}}^{(1)}\right)Z_0,\\\nonumber
	Z_3=&\,\left(\frac{1}{6}\tau_2^3\mathscr{D}_{\nu}^{(3)}\mathscr{D}_{\bar{\tau}}^{(3)}  -\frac{i\pi}{2}\tau_2\nu\mathscr{D}_\nu^{(2)}\mathscr{D}_{\bar{\tau}}^{(2)}-\frac{3i\pi}{4}\tau_2\mathscr{D}_{\nu}^{(1)}\mathscr{D}_{\bar{\tau}}^{(2)}
	-\frac{\pi^2}{4\tau_2}\nu^2\mathscr{D}_{\nu}^{(1)}\mathscr{D}_{\bar{\tau}}^{(1)}  -\frac{\pi^2}{2\tau_2}\nu\mathscr{D}_{\bar{\tau}}^{(1)}\right)Z_0.
\end{align}

\par \vspace*{.5cm}\par

%
\bibliography{refsJTb}
\bibliographystyle{bibstyle2017}

\end{document}